\documentclass[11pt,a4paper]{article}
\usepackage[T1]{fontenc}
\usepackage[utf8]{inputenc}
\usepackage[pdftex]{graphicx}

\usepackage{pdfpages}

\usepackage{pslatex}
\usepackage{times}

\usepackage[pdftex,colorlinks,urlcolor=blue]{hyperref}

\begin{document}

\vfill{}
\title{Phylogenetic Tools in Astrophysics}
\vfill{}

\date{\small Feb. 2017 Wiley StatsRef: Statistics Reference Online \\ DOI: 10.1002/9781118445112.stat07935}

\author{\small Didier Fraix-Burnet \\ \small Univ. Grenoble Alpes, CNRS, IPAG, F-38000 Grenoble France}

\maketitle

\textbf{keywords} %5 to 10
Astrophysics - Multivariate Analyzes - Cladistics - Maximum Parsimony - Unsupervized Classification - Clustering

\textbf{Abstract} % (150-200 mots)
Multivariate clustering in astrophysics is a recent development justified by the bigger and bigger surveys of the sky. The phylogenetic approach is probably the most unexpected technique that has appeared for the unsupervised classification of galaxies, stellar populations or globular clusters. On one side, this is a somewhat natural way of classifying astrophysical entities which are all evolving objects. On the other side, several conceptual and practical difficulties arize, such as the hierarchical representation of the astrophysical diversity, the continuous nature of the parameters, and the adequation of the result to the usual practice for the physical interpretation. Most of these have now been solved through the studies of limited samples of stellar clusters and galaxies. Up to now, only the Maximum Parsimony (cladistics) has been used since it is the simplest and most general phylogenetic technique. Probabilistic and network approaches are obvious extensions that should be explored in the future.

\section{Introduction}

Extragalactic astronomy is a good illustration of the new era that is opening. Galaxies were discovered by Hubble one century ago, and the only data he had was images at visible wavelengths. He classified the morphologies of his observed galaxies by eye into four classes, and a few years later, from a physical argument, he depicted the evolutionary relationships between three of them on the famous Hubble Tuning Fork Diagram. This classification, slighly refined, is still made by eye and used as a general classification of galaxies despite the fact that the technology provides us with much more information (different wavelengths, spectra) on millions and millions of objects. 
However, a few multivariate unsupervized classification (\href{http://dx.doi.org/10.1002/9781118445112.stat06490.pub2}{stat06490.pub2}, \href{http://dx.doi.org/10.1002/9781118445112.stat05127.pub2}{stat05127.pub2}, \href{http://dx.doi.org/10.1002/9781118445112.stat05612}{stat05612}) studies have begun to appear in the extragalactic literature at the turn of the XXIst century$^{\cite{Fraix-Burnet2015}}$. The phylogenetic approach is one of them$^{\cite{FCD06,jc1,jc2}}$.

Phylogenetic tools are heavily used in evolutionary biology (\href{http://dx.doi.org/10.1002/9781118445112.stat05446.pub2}{stat05446.pub2}, \href{http://dx.doi.org/10.1002/9781118445112.stat07642}{stat07642}), even though linguists (\href{http://dx.doi.org/10.1002/9781118445112.stat00088}{stat00088}) have probably invented the principles of the methodology. The idea is strongly based on the darwinian process of transmission with modification: two entities are said to be close if they share a same trait inherited from a common ancestor. This explains the hierarchical organization of the diversity of living organisms. The simplest and most general technique to build a phylogenetic tree is cladistics (\href{http://dx.doi.org/10.1002/9781118445112.stat05363}{stat05363}), also called Maximum Parsimony (\href{http://dx.doi.org/10.1002/9781118445112.stat01733}{stat01733}, \href{http://dx.doi.org/10.1002/9781118445112.stat05770}{stat05770}). 

In unsupervized classification, the most common approach is to aggregate objects into clusters. 
The origin of these clusters is better understood through their relationships. The phylogenetic approach conversely looks for these relationships and then defines families and sub-families depending on the inheritance scheme depicted by the tree$^{\cite{Fraix-BurnetHouches2016}}$ (\href{http://dx.doi.org/10.1002/9781118445112.stat06461}{stat06461}).

Why and how is it possible to use the phylogenetic approach in astrophysics where the diversification process is not comparable to sexuation or duplication? There are at least two answers to this question.

Firstly, the role of evolution in the tree-like representation provided by cladistics is debated$^{\cite{Pecaud2014}}$. For instance, is darwinian evolution really a prerequisite for cladistics, or are the relationships interpretable in terms of evolution? Indeed, this may not be so important in practice, since the key point for the hierarchical structure is the transmission with modification as illustrated by the success of the phylogenetic approach in linguistics. Very generally, the evolution of any single object can be seen as a transmission with modification process. However, the evolution of an ensemble of objects is not necessarily represented by a hierarchical tree if there are too many parallel or regressive evolution of some properties, or hybridization as for bacteria. In any case, a phylogenetic tree depict the shortest path to transform an object into another one, and this transformation can be virtual, like in the example of volcanoes$^{\cite{Hone2007}}$.

Secondly, all astrophysical objects evolve, and we can nearly always identify a transmission with modification process. For instance, stars form from clouds of gas. This gas is enriched into heavier chemical elements inside the more massive stars, which then explode and eject this new gas that will later form new stars. For galaxies, when two galaxies merge, the new object inherits from all the material of its progenitors (stars, gas and dust) but with many modifications (formation of new stars, modified kinematics, and global distinct chemical composition).

Do we expect a hierarchical organization of the diversity of astrophysical entities? This is a priori difficult to guarantee, but the increase of the metallicity of stars and gas due to the previous generations, and the complexity of the transformation processes for galaxies that very probably induced an increase of diversity with time, seems to open this possibility. In any case, nothing prevents the use of phylogenetic algorithms on astrophysical data as exploratory tools. And this has now been done on several kinds of astrophysical objects, with great success.

\section{Phylogenetic Algorithms} 

The most natural phylogenetic algorithm is cladistics, or Maximum Parsimony (\href{http://dx.doi.org/10.1002/9781118445112.stat05363}{stat05363}). It can be readily applied to any kind of data since there are a very minimal assumptions. The idea is to look for all the possible arrangements of the objects in study on a tree, and select the most parsimonious one as the best evolutionary scenario. The objects are called taxa which can be individuals or species.

The parsimony criterion is based on the total evolutionary cost. To go from an object to another on the tree, some properties must be changed by certain amounts. The total cost of the tree is the sum of all these changes which are called steps. 

Let us illustrate this point with the Hubble Tuning Fork Diagam mentioned above. Let us consider two parameters that describe the rough morphology of galaxies: the presence/absence of a disk, the presence/absence of a bar. Elliptical galaxies have no disk and no bar (0,0), spiral ones have a disk and no bar (0,1) and barred spiral ones have both (1,1). Since there is only one possible arrangement on a tree with three objects, the simplest scenario is to put one change for the disk property on the branch connected to the ellipcal galaxies, and one change for the bar property on the branch for the barred spiral galaxies. According to Hubble, the elliptical galaxies should flatten in a disk with time. Hence the absence of a disk (value 0) is an ancestral state and its presence (value 1) a derived state acquired by the two groups of spiral galaxies$^{\cite{Fraix-BurnetHouches2016}}$.

To perform the computation of the number of steps of a tree, discrete data are necessary, each discrete value is called an evolutionary state. The parameters to use are called characters and are the ones that can be given at least two evolutionary states, one being called ancestral and the other one derived. The parameters are observables or properties describing the objects, the characters being supposed to keep a trace of the history. In astrophysics, all quantities are quantitative and continuous, so that discretization is required. This is a complicated subject in statistics (see stat02713), but we have found that about 20 to 30 equal bins is a good compromize between objectivity, stability and the continuous nature of the parameters. 

Some additional hypothesis can be imposed to the cost between two evolutionary states. Probably the most general one is the Wagner optimization criterion which is a l1-norm and allows for reversals. But irreversibility can be imposed or any more or less complicated hypothesis through a cost matrix.

A rather unique feature of the cladistics algorithm is its ability to take into account uncertainties and even missing values (\href{http://dx.doi.org/10.1002/9781118445112.stat03968}{stat03968}). By providing a range of possible states instead of a single one, the Maximum Parsimony procedure selects the state which minimizes the total evolutionary cost, value which can be regarded as a prediction. This feature can be quite appreciable in astrophysics where uncertainties are omnipresent.

At this point it should be clear that the minimization process of Maximum Parsimony is very similar to the Mimimum Spanning Tree technique  (\href{http://dx.doi.org/10.1002/9781118445112.stat06487}{stat06487}) with the l1-norm. The big difference is that in a phylogeny, internal nodes are added and represent hypothetical ancestors. Formally, it is postulated that the ancestors be unknown to allow for future discoveries. These internal nodes widen considerably the possible topologies of the trees, at the expense of the computing time (NP-hard problem).

To alleviate this computing limitation of cladistics, other methods have been developed. Among the other character-based approaches, the Maximum Likelihood algorithm is quite popular in biology. It is a probabilistic method which selects the most probable tree according to some evolutionary model$^{\cite{Williams2003}}$. This technique has not yet been applied in astrophysics since it requires a good knowledge of the evolutions for the characters. However this should be certainly investigated in the future.

The other big category of methods uses pairwize distances between the taxa (\href{http://dx.doi.org/10.1002/9781118445112.stat02470}{stat02470}, \href{http://dx.doi.org/10.1002/9781118445112.stat02231}{stat02231}). These distances are computed from the characters and the algorithms are generally extremely fast and efficient. It has always been somewhat surprising that character- and distance-based approaches  yield similar results since the mathematical connection is still not obvious$^{\cite{TF09,TF15}}$. This might be due at least partly to the fact that the characters have an evolutionary information (the phylogenetic signal) that is not entirely lost when computing distances. The most popular algorithm is the Neighbor Joining Tree Estimation$^{\cite{NJ1987,NJ2006}}$ which is a bottom-up hierarchical clustering method (\href{http://dx.doi.org/10.1002/9781118445112.stat02449.pub2}{stat02449.pub2}). It would sound like an interesting choice in astrophysics, however it is not easy to assess the robustness of the result tree without comparing with cladistics analyzes.

\section{Applications}

Up to now, the Maximum Parsimony method has been used for galaxies$^{\cite{FCD06,Fraix2010,Fraix2012}}$, globular clusters$^{\cite{FDC09}}$, stellar populations$^{\cite{FD15}}$, Gamma Ray Bursts$^{\cite{Cardone2013}}$ and the Jovian satellites$^{\cite{Holt2016}}$. 
In all cases, cladistics provides the identification of families of objects that are characterized by an interpretation in terms of formation and evolution that could not have been discovered otherwize. For instance, the analysis finds three families of Globular Clusters of our Galaxy together with their evolutionary relationships. The average properties of each family is found to be specific of a certain formation environment, such as the chemical content of the gas and its dynamics. The three families thus correspond to three stages in the assembly of our Galaxy$^{\cite{FDC09}}$.

The phylogenetic approach in astrophysics was initially thought for galaxies because of their intrinsic complexity and the complexity of the diversification throughout the evolution of the Universe$^{\cite{jc1,jc2}}$. The demonstration of this idea was given$^{\cite{Fraix2012}}$ with the first evolutionary tree showing how the transformation processes of galaxies shaped their diversity. This was also the first multivariate classification of galaxies, even though it is based on a limited sample of less than one thousand individuals. Ongoing studies are tackling bigger samples.

There is a fundamental difficulty in phylogenetic approaches, that is indeed true also for any multivariate classification, and biologists have been struggling for several centuries for this problem: which set of parameters should be used for multivariate classification (\href{http://dx.doi.org/10.1002/9781118445112.stat05615}{stat05615})? It may depend on the goal of the classification, but if the understanding of the entire diversity and its origin is desired, then an objective selection of the parameters is required. Taking  blindly all available observables or properties would probably lead to a negative result, either because of the curse of dimensionality that smears out structures in the parameter space (\href{http://dx.doi.org/10.1002/9781118445112.stat00408}{stat00408}), or because of dominant or conflicting parameters. 

Phylogenetic approaches are particularly sensitive to the quality of the characters since they should be the tracers of the relationships between the species. Synapomorphies are characters for which innovations are transmitted to all descendant branches of a given node (the hypothetical ancestor) and are maximized by the Maximum Parsimony procedure. Conversely the homoplasies are minimized since they tend to destroy this transmission behavior and consequently the hierarchical structure. Reversals, convergent and parallel evolutions are homoplasies.

Despite some mathematical rules that would probably help$^{\cite{TF09}}$, it remains difficult to find the best parameter set without trial-and-error computations. Statistical analyzes like Principal Component Analysis (\href{http://dx.doi.org/10.1002/9781118445112.stat06472}{stat06472}) give some insight on the parameters and their correlations, but it may not always be obvious to distinguish a size effect from a confounding correlation (\href{http://dx.doi.org/10.1002/9781118445112.stat06786}{stat06786},  \href{http://dx.doi.org/10.1002/9781118445112.stat06194}{stat06194}) due to evolution$^{\cite{DFB2011}}$. The first one is probably not very informative while we are clearly looking for the second one. For illustration, a cladistic analysis reveals that the origin of the correlation in galaxies between the central velocity dispersion and the metallicity is certainly evolutionary, i.e. the two properties independantly and statistically evolve monotically$^{\cite{Fraix2012}}$. Other statistical investigations can be performed, and with the multiplication of phylogenetic studies, a better idea of the parameters to use will emerge.

\section{The Future}

The application of phylogenetic methods is rather recent in astrophysics, and many conceptual, technical and philosophical questions had to be first addressed. We are however now in a position to use these approaches more widely, explore, adapt or develop new algorithms which can be specifically more efficient for the analyzes of astrophysical entities.

Maximum Parsimony will probably remain a reference because of its conceptual and practical relative simplicity to implement. Its main limitation is to be a NP-hard problem. We have to build the phylogenies piece by piece, much like the Tree of Life. Other approaches, like distance-based phylogenetic methods or multivariate clustering techniques, could probably be used in parallel to build classes or sub-classes before finding their relationships with Maximum Parsimony.

Probabilistic methods like the Maximum Likelihood is also promising since the basic physics of the astronomical objects and their evolution is in principle well known. Their complexity prevents precize prediction through heavy numerical simulations, but it seems reasonable to think that some probabilistic laws could be found at least for some characters. Also bayesian phylogenetic methods would deserve some attention in the future.

The network (\href{http://dx.doi.org/10.1002/9781118445112.stat02300}{stat02300}) representations of the diversification of the astrophysical entities is a promising direction. It has now been shown that the tree-like representation of different species of astrophysical objects is physically reasonable, but it appears, especially for galaxies, that a more complex scheme could be required$^{\cite{Fraix-Burnet2015}}$. 
Firstly, a network representation can synthetize conflicting trees$^{\cite{Huson2006}}$ in a more informative way than consensus trees. These conflicts may arize from a lack of information on the population of objects, which is probably often, if not always, the case. Secondly, hybridization, or parallel or convergent evolution of some characters, certainly occur in galaxy evolution. This is known to destroy the tree-like structure by creating perpendicular branches, making what biologists call reticulograms. In any case, networks are a more general representation of evolutionary relationships than trees. They are also more difficult to read and to use for a physical interpretation.

\section{Related Articles}

 Statistics in Astrophysics, 2014, E. Feigelson, \href{http://dx.doi.org/10.1002/9781118445112.stat00155}{stat00155}

 Ultrametric trees, 2014, F. Murtagh, \href{http://dx.doi.org/10.1002/9781118445112.stat06461}{stat06461}

% Systematics, numerical methods (http://dx.doi.org/10.1002/9781118445112.stat07642/abstract)
%stat07642

% Classification, Overview (http://dx.doi.org/10.1002/9781118445112.stat05612/abstract)
%stat05612

% % In-line references given in the text
% Cladistics (http://dx.doi.org/10.1002/9781118445112.stat05363/abstract)
%stat05363

% Binning (http://dx.doi.org/10.1002/9781118445112.stat02713/abstract)
%stat02713

\end{document}